# The synchronic principle for a new scientific method

Paolo Silvestrini

Dipartimento di Matematica e Fisica, Università della Campania "Luigi Vanvitelli", Viale Abramo Lincoln 5 , 81100  Caserta, Italy

and

Istituto di Scienze Applicate e Sistemi Intelligenti ISASI-CNR, Comprensorio A. Olivetti Edificio 70, Via Campi Flegrei 34, 80078 Pozzuoli , Napoli, Italy

**Abstract**: Local realism has been superseded by recent theoretical and experimental developments in quantum physics which show a synchronic connection between the parts. In contrast to the principle of local realism, which is the basis of classical reductionism, I have called the principle of non-locality at the basis of a new method of scientific investigation "synchronic principle". The synchronic principle introduces a new element in the approach to reality, which with anthropomorphic terms we could call "conscious choice". In a synchronic paradigm, evolution does not occur in a deterministic way, but is characterized by a certain intrinsic freedom of choice that we associate with awareness. Considerations based on the evolution of the universe towards the emergence of life lead us to think that this "awareness" also exists in the vacuum state: in the model I proposed a definition of the agents of consciousness in terms of arrays of correlated qu-bit, from the simplest consisting of 6 qu-bits of the quantum vacuum (which result as a superposition of 64 base states), up to the most complex organisms consisting of a very large number of correlated quantum bits. Some of the fundamental vacuum states of consciousness can be related to known elementary particles (leptons and quarks) and to fundamental interactions. The rest constitute purely non-local elements of consciousness with no counterpart in terms of wave or particle within the local observation limit. In this model, consciousness processes quantum information through logic gates and collapses as a result of observations, like a quantum computer does. Synchronicity does not in deny the conclusions reached by traditional science can be used at the local level, but opens scientific knowledge to new possibilities, and invites us to look at the world through a different, not local awareness.

## 1.Introduction

The phenomenon of quantum coherence and entanglement in macroscopic systems is supported both by theoretical consideration (Caldeira 1981; Zurek 1981; Zurek 2003) and by a large amount of experiments (it is impossible to mention all the experiments: see for instance Arute 2019; Chiorescu 2003; Emary 2014; Knee 2016; Martinis 1987; Nakamura 1997; Rouse 1995; Silvestrini 1997; Wan 2016) opening the doors to scientific speculations of great scope, both from the applicative point of view and from the philosophical and ethical one. I think it is important that humanity as a whole, and not just a narrow scientific oligarchy, becomes aware of the possibilities that are opening up with these new discoveries and the new worldview connected to them (Bohm 1980; Leggett 2004; Nielsen 2000).

Recent developments in quantum physics undermine one of the most radical founding principles of the classical scientific method (I am referring to experiments on the so called EPR-Bell inequalities: see for instance  Aspect 1981; Freedman 1972; Giustina 2015; Groeblaker 2007 Hansen 2015; just to cite some of the them), called the local realism principle, and therefore require a profound revision of the scientific method by extending it to phenomena not directly measurable at the local level.

I would like to clarify what I mean by classical locality and, as opposed, by quantum non-locality, which I also called "synchronicity": locality is a way of conceiving physics that attributes reality only to physical systems that can be defined through the characteristics that can be measured in the immediate vicinity of the system itself. Any other distant object cannot contribute to the reality of the local system. The interactions are also local, the presence of which can be measured in the immediate vicinity of the systems under consideration. In a local and mechanistic vision, reality is always separable into elementary parts, possibly interacting more or less intensely. This is the basis of classical scientific reductionism.

Non-locality, on the other hand, indicates that distant physical systems can be related in an inseparable way, to form a single quantum system. This correlation is not measurable at the local level, but means that the overall system cannot be separated into simple elements without losing its essential characteristics. Entanglement and quantum synchronicity are non-local phenomena, of a "mysterious" nature because they cannot be traced back to actions related to the cause-effect principle (Bohm 1980).

As opposed to classical reductionism, I would like to call the principle of non-locality at the basis of a new method of scientific investigation "synchronic principle", and the new science that derives from it "Synchronic Science". Quantum Physics, while having shown the need to introduce synchronicity in our vision of nature, has remained in its interpretation strongly linked to classical reductionism, confining its attention to that limited subset of phenomena that are verifiable in local experiments. Of the phenomena that are based on the complexity of the synchronic principle, not only does official science ignore them, as they cannot be verified in reductionist experiments, but tends to label them as on the border between science and pure philosophical speculation, or worse as "magic" from sorcerer or scientific heresy. An exception are the countless studies of the last two or three decades on entanglement and macroscopic quantum coherence, with enormous funding worldwide for research on quantum technologies; the motivation for this widespread interest, however, is not to be found in the awareness of the need for the synchronic principle to explain complex natural phenomena, but rather in the economic and military interest for the development of the quantum computer that is made possible by these studies.

The synchronic principle introduces a new, I would say revolutionary, element in the approach to complexity, linked to the possibility of "non-mechanistic choice" that complex systems can have in their evolution. This possibility, which with anthropomorphic terms we could call "conscious choice", is present in correlated systems through quantum synchronicity, and could be much more generally widespread than we can hypothesize in a reductionist paradigm. In a synchronic view, consciousness could be a fundamental ontological element, and it would therefore make no sense to ask how it emerges in reductionist terms, or why it exists. It just exists. We can only ascertain its presence and accept its indubitable factuality, and try to understand its ways of manifesting and interacting.

This is one of the fundamental research topics I envision for synchronic science. In a synchronic view, the elementary particles are also simple elements of consciousness, which self-organize to form complex organic systems endowed with an articulated and evolved consciousness, whose greater degree of synchronic integration of the simple elements leads to responses to the increasingly free and responsible stimuli (Hoffman 2008).

In this article I propose a synchronic model for conscious agents starting from the fundamental state of the universe (the quantum vacuum): this model could be the basis for a unified model of subject and object, materiality and awareness. As it is constructed, it allows an association of some elements of consciousness with the elementary particles interacting through the fundamental fields, so as to automatically lead back to quantum physics in the simplest cases. In the more general case, it contains an enormous variety of overlapping possibilities, of which only some can be associated with the known matter; others are instead peculiar states of consciousness that interact according to the modalities of quantum computation.

This article is intended to be an experiment of how a "new" scientific method can be imagined, which integrates rigor and imagination, in a knowledge that is both objective and subjective at the same time. The invitation is to stimulate a discussion on this aspect, and I absolutely do not pretend that my considerations are in any way definitive or even just guidelines valid for everyone.

## 2. A conscious evolution

The process of evolution has gone through a very large number of bifurcation points in which parameters barely different from those observed would have led to absolutely different and lifeless evolutions. This is true right from the choice of the universal constants that determine the mass and charge of elementary particles and the relationships between the known force fields, which as far as we know could have been different from those observed, up to the very delicate initial conditions that led the highly chaotic evolution of the universe to the present remarkably ordered manifestation - in terms of clusters, galaxies, stars, planets, up to and including biological organisms - in the approximately 15 billion years that we estimate have elapsed since the big bang. The development of life on earth is a "miracle" of perfect harmony of conditions so exponentially improbable that it is really difficult to think that it was determined solely by chance.

To express this extremely improbable consideration, Carter introduced the anthropic principle in 1973.[1] He proposed two versions of the principle: first a "weak" Anthropic Principle in consideration of the fact that "our position in the universe is necessarily privileged insofar as it must be compatible with our existence as observers ". He also proposed a "strong" Anthropic Principle, which states: "The universe (and therefore the fundamental parameters on which it depends) must have those properties that allow conscious observers to develop within it at a certain point in its history."

The anthropic principle wants to underline that we live in a universe that allows the existence of life as we know it; in a local view this fact is so particular that if we tried to mathematically calculate the probability that evolution will lead to the birth of life through completely random processes, this probability would be zero. The universe is a complex system, whose chaotic evolution passes through different phases of "bifurcation", in which imperceptible differences in boundary conditions determine completely different developments. The probability that many critical parameters randomly assumed the exact values necessary for the development of life is zero. So it can be assumed that the emergence of consciousness is not the result of chance.

If life cannot be just the result of chance, does this mean that there is an end in the evolution of the universe? Does an intelligence guide evolution so that life manifests itself? Do we have to admit a principle that states that the presence of a conscious life is essential for the very existence of the universe?

Now the request may seem excessive: after all this is the only universe we have experience, which contains man and life. We know of no others to refer to for a hypothetical probability calculation.

---

[1] The term "anthropic principle" was coined in 1973 by Brandon Carter in his speech "Large Number Coincidences and the Anthropic Principle in Cosmology" during a symposium held in Krakow as part of the celebrations for the 500th anniversary of the birth of Copernicus. In fact I think it can be said very plausibly that the idea of the anthropic principle (although obviously not under this name) was actually introduced by Boltzmann about 100 years earlier, pointing out that it is only in an "atypical" region of the Universe that human life has been able to develop.

However, the above questions are legitimate and I would like to try to prove their validity with a simple example.

Let's consider the Rubik's cube for this purpose: in its original version, it has 9 squares on each of its 6 faces. The squares differ in color, with a total of 6 different colors. When the Rubik's cube is solved, each face has all nine squares of the same color. It starts with a random configuration, with different colors on each of the faces. The aim of the game is to trace the original position of the small cubes in each face by bringing the whole cube to have a single color for each face.

The cube can take on a huge number of possible combinations of which only one is the correct one. This number can be calculated and turns out to be $N = 43\ 252\ 003\ 274\ 489\ 856\ 000$.

Now let's imagine observing the evolution of a Rubik's cube, whose configurations have changed by "invisible" hands. In the event that the changes occur randomly, the probability of reaching the solution is very low: supposing to see the configuration change a hundred times every second, it would still be necessary, to solve the cube, a time equal to approximately the estimated age of the universe.

We realize that in reality there is a purpose in the game from the fact that the configuration changes, although they may initially appear random, are actually not such: we will see favored combinations that aggregate cubes of the same color on the same face. If the intelligence that moves the "invisible" hands is an expert in the game, the solution will be quickly reached. It could even be possible to solve the cube in the shortest possible time by making as few changes as possible. In this case, no movement is casual, but each one has a specific purpose towards the final purpose. The world record for the solution time is in fact less than 5 seconds. Even if the intelligence is not an expert in the game - and will therefore move with a certain randomness - observing the result of the movements will lead to preserving the configurations that aggregate cubes of the same color on each face, thus directing towards the solution, which will take place in a time not very short but still reasonable, much faster than with completely random changes.

Now let's consider the fact that, if we don't see invisible hands and watch evolution happen "naturally", we could naively think that there is no intelligence guiding the movements and therefore believe that the evolution we observe is the only possible one, simply because we don't move the cube ourselves. We could strive to discover the regularities in the movements of the cube configurations, and of course we would find them. We could even summarize them in simple mathematical equations and call them "physical laws", which will have a more deterministic character the more expert is the intelligence that guides evolution; paradoxically, once we have found those laws, we can think that we are particularly "intelligent" and rational for having discovered them.

In reality, looking at the phenomenon from a broader perspective, the discovery of those physical laws actually indicates a certain rationality and logic in highlighting the regularities of movements, but at the same time highlights a blindness that prevents us from seeing the hands, and to therefore recognize the true intelligence that guides the process.

A greater capacity for understanding could instead recognize in those same laws the presence of invisible hands and a purpose in the game, thus implicitly admitting one's own blindness.

Let's go back to the universe in which we find ourselves, which is obviously an immensely more complex system than the Rubik's cube: here we observe an evolution that led to the appearance of life and our presence. This configuration is so critical and so enormously unlikely that it requires the presence of intelligence and a specific purpose in this sense.

This request contains the essence of the anthropic principle, at least in its strong formulation, which perhaps could be more properly called the "biocentric" principle, as it places consciousness as the ultimate purpose of existence.

The synchronic principle, which has been observed in countless experiments and which is part of the formalism of Quantum Physics can replace the anthropic principle, which like all ad hoc principles leaves even the most staunch defenders of reductionism unsatisfied. In fact, synchronicity manifests itself to a local vision as significant coincidences without there being a cause-effect relationship between the events considered. Events correlated through entanglement do not occur by chance, but they are part of a single quantum process that cannot be separated (either in space or time) into distant parts. The universe that presents itself to us in the quantum model is a non-local, intrinsically connected universe, in which any separation into locally defined parts in general is not possible: every observer is entangled with what he observes and the universe is a single system inconceivable in any classical model (Bohm 1980).

On the basis of this synchronic principle we have hypothesized a widespread possibility of non-random choice, which with an anthropomorphic term we can call consciousness or awareness, as long as we are careful not to identify this awareness with human consciousness: in fact this universal awareness is present at every level, from elementary particles to complex systems capable of self-organizing in an order of articulation that is increasing inclusive to eventually include the whole universe (Fields 2016).

### 3. Elementary particles of consciousness

The main challenge of a synchronic science is to build within the formalism of quantum mechanics an "observer", that is, an entity whose states correspond to a reasonable model of conscious awareness. What defines a conscious agent? how does it operate and how does it evolve in the universe? what is a conscious observer and a non-local observation? how can quantum information be processed without reducing it to a set of classical bits?

The new science must try to gear up to answer these questions (Fields 2018, Hoffman 2008).

Let us now try to define a "quantum observer" that can serve as a fundamental paradigm of synchronic science, on the basis of which we can build a theory that can be traced back to quantum physics in the limit cases of local observations.

In quantum physics, any subset of the universe can be considered an observer, and therefore, by accepting the synchronic principle, a more or less complex element of consciousness. From elementary particles to systems with ever increasing complexity, consciousness must be able to expand and evolve in an increasingly inclusive way: simple elements of consciousness, which we can call elementary points of view, combining themselves must be able to form elements of consciousness that are inclusive of the sum of the original points of view, but in general also capable of a broader, non-local observation, not attributable to the simple sum of the parts. A definition of an element of consciousness must therefore satisfy the condition that the union of several elements is still an element of consciousness capable of a complexity of observation greater than or equal to the sum of the individual constituents. Furthermore, the elements of consciousness must be connected with a certain level of synchronicity, and therefore never completely separable. The separation, albeit arbitrary, can be done by defining fictitious borders between the parts of the universe; but in this case it is necessary to hypothesize an exchange of information and the possibility of the collapse of the wave function that we encountered in quantum physics.

We begin by analyzing the definition of conscious agent as proposed by Hofmann and coauthors (Hoffman 2008; Fields 2018) through the mathematical concept of a classical Markovian kernel. In this proposal, the conscious agent is in contact with the world (defined as that part of the universe other than itself) and interacts with it through a circular connection of three processes, identified as characterizing consciousness: perception, decision and action. The conscious agent has experiences through interaction with the world, on the basis of these perceptions he decides what actions to take, and as a result of these actions he influences the state of the world; and the process resumes in a circular connection of cause-effect relationships: the observer appreciably perturbs the characteristics and evolution of the observed system; and for its part, the observed system strongly influences the behavior of the observer, and the same rules and methods of observation. This interaction, with its characteristics of circularity, is amplified more and more when the observer's perception capacity widens, who is subject to unpredictable evolutions as a result of the stresses to which he is subjected.

Although this definition provides for a strong interaction between the observer and the observed world, the approach is still one based on reductionism.

The reductionist and causal aspect of the definition of conscious agents by Hofmann and co-authors is formally recognizable in the way in which the conditional probabilities underlying their definition of conscious agent are mathematically treated, since they are considered independent: for example if the state of the world is $w_1$, then the set of possible conscious experiences that could result is described by a list of independent probabilities, with the only condition that the sum is 1; if the state of the world were $w_2$ then the mixture of probabilities would be different, but still independent, not coherently superimposed. This is the way to treat statistics in a classical view, that is, as a mixture of probabilities, with the only condition that the sum must be 1, that is certainty, if all the possibilities are taken into account. In a quantum view, the possibilities are not independent, either one or the other according to a probabilistic law, but the possible alternatives are coherently superimposed, with the fundamental implication of interference effects. The same classical reductionist approach is also formally recognizable in the description of the observer (conscious agent) and of the observed (the world). Despite the symmetry of the definition, whereby an observer can become the observed and vice versa in an exchange of roles that is always possible and legitimate, however the states that describe the two are considered separate and not entangled. The interaction between the two is of a cause-effect type, albeit in an inextricable and in fact inseparable circularity given the complexity of the interaction. This type of correlation produces an evolution that needs time to manifest itself, in a circular cause-effect sequence that we cannot know in detail, given its complexity, but which in fact we admit exists in principle.

Synchronicity is a completely different type of connection, which connects the systems involved in an a-causal, inseparable, synchronic way regardless of their cause-effect relationships, regardless of their space-time distance. Synchronicity is an ever-present connection, which does not need time or causes to manifest itself: it exists and is an ontologically primary part of the universe, even if it is not visible to our objective perception.

The first thing I would like to try is to extend Hofmann's definition by replacing the classical probability mixtures that he proposes with a coherent superposition of possibilities according to quantum formalism, and at the same time to introduce the possibility of entanglement and collapse of the wave function. This too can be done within the mathematical formalism of quantum mechanics.

In a new synchronic scientific method, each element of reality must intrinsically contain the essential characteristics that we attribute to consciousness: exchange of information with one's world (perception), a certain freedom of choice (decision), freedom to act in the decided direction (non-random action). It is through these communication channels between the conscious agent and the world with which it is entangled that the evolution towards universal awareness is determined.

We should start our hypothesis from the simplest possible case, and try to understand how states of consciousness could be applied to known elementary particles. If we succeeded in this analogy, then the correspondence with quantum physics in the local and causal approximation would be automatic, while the synchronic theory could give indications of new implications.

The most elementary state possible is the fundamental state of the Universe, that is, the state of vacuum. In field theory, the quantum vacuum is not static, but a highly dynamic state in which virtual elementary particles are continuously produced, whose average life is too short to be observed. The vacuum state is energetic and all virtual particles are continuously created and destroyed, compatibly with the conservation of fundamental quantum numbers. In the quantum vacuum all the fields are present, albeit in the state of minimum energy, and all the virtual particles that are the elementary excitations of these fields, in an inextricable synchronic intertwining. Let's try to build a model of synchronic elements of consciousness for the quantum vacuum, from which the most complex conscious agents can emerge through a process of self-determination, just as in a causal vision the aggregate matter emerges from the elementary particles through a causal evolution.

The three characteristics of perception, decision and action must also be present at the fundamental level. In the elementary case, each of these intrinsic degrees of freedom can have the minimum number of possible states, that is, two: we will have two possible states of perception, two possible decision states and two possible basic actions. The combinations of all possible base states are $2^3$, that is 8. In a synchronic view, the observer's elementary state of consciousness can exist in any combination of coherent superposition of all 8 base states. Together with the elementary observer, and symmetrical to it, the elementary observed system must exist, complementary to the first: this system constitutes the world of the observer, but at the same time it can be considered as the observer by reversing the roles: the two systems must be entangled. The basic states of the observer-observed entangled system are therefore 64, that is, all the possible combinations of the 8 states of the first with the 8 of the second, in a perfect symmetry of interchange of roles. The state of consciousness of the entangled elementary system will therefore be any coherent superposition of all 64 base states. Since the possible superposition states are infinite, already at the elementary level of consciousness the information contained in any state cannot be represented by a sequence (finite or infinite) of classical bits[24]: however it can be represented by an array of 6 qubits entangled, using the concepts of quantum information. In fact, I believe that the consciousness of complex conscious systems is a self-organized dynamic system capable of processing quantum information, a bit like we are trying to do with technologies related to the development of quantum computer.

According to the logical scheme that I have set, I would now like to try to associate the basic states of elementary consciousness with the elementary excitations of the vacuum state of quantum physics, that is, the elementary particles. Of course, not all 64 states of elementary consciousness must necessarily associate with known elementary particles; some may be forms of consciousness that cannot be traced back to measurable objective reality, but it is worth making general assessments. This association is generic and highly speculative, but it still seems useful to show a certain reasonableness, since any theory that starts from the definition of the agents of consciousness must be traced back to quantum physics in the limiting case of predominance of the causality principle. First of all, let's try to take into account the perfect symmetry between observer and observed: to visualize the argument I am about to do, let's establish to represent the 8 basic states of the observer and the observed through trigrams (ordered sequences of three superimposed lines, which can be whole or broken), in which the two states of perception, decision and action are represented by continuous or broken lines placed one above the other. The 8 possible states of the observer are therefore those shown in the following diagram (drawn in black), and we can do the same for the observed system (in blue).

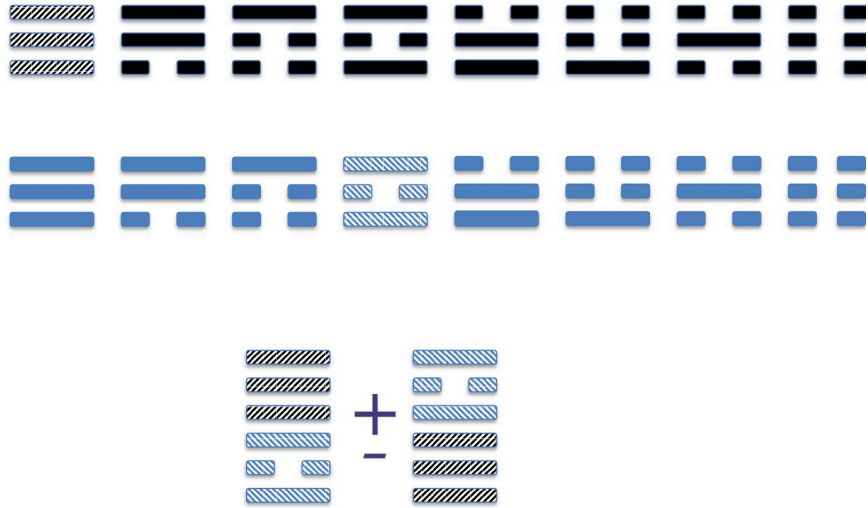

The combinations of the entangled system will be represented in this schematization by the 64 hexagrams (sequences of six whole or broken lines) which are obtained by combining the 8 upper trigrams (relative to the observer) with the lower 8 (relative to the observed), of type the two ones that are drawn in the diagram, as obtained by combining the first with the fourth. Due to the perfect symmetry between the observed system and the observer, the two states obtained by inverting the order of the trigrams must correspond to equivalent global situations. In quantum mechanics, in the presence of equivalent states, stationary states are coherent superpositions of the two symmetrical states that are constructed by adding and subtracting the two states (the two superpositions are called symmetrical and anti-symmetrical), as shown in the figure. These overlaps can be constructed only if the trigrams that form the hexagram are different (which happens for 56 hexagrams), while in the 8 hexagrams made up of identical trigrams we have only one possible combination. The symmetrical overlap is generally more stable, while the antisymmetric combination needs more energy to be activated.

Ultimately, of the 64 possible overlapping basic states to form elementary states of consciousness we have 8 hexagrams made up of two identical trigrams (which correspond to an equal state of consciousness between observer and observed), 28 symmetrical elementary superimpositions, and 28 antisymmetric superimpositions of hexagrams consisting of different trigrams with inversion between observer and observed. Therefore, using the formalism of quantum mechanics to define the elementary states of consciousness of the vacuum, we have 28 states that are obtained by breaking the symmetry, plus 8 consisting of identical trigrams, for a total of 36 states. In addition to these we have 28 antisymmetric combinations that can be activated with greater difficulty. The vacuum state of consciousness will then be any coherent superposition of these basic states, which can be described as an array of six quantum bits entangled, in a representation proper of quantum information.

It should be noted that in a local view, introducing the collapse of the wave function, from this enormous potentiality of possible states we will always obtain only one of the 64 possible different manifestations, with different probabilities of occurrence linked to the degree of stability of the relative base states.[25]

In a synchronic view, of course, the paradigm is completely different, as the interpretation we must give to evolution: this is because evolution does not occur over time according to causal processes described by physical laws, but it is a synchronic manifestation of universal awareness, whose beauty and harmony can be represented locally by physical laws.

In this hypothesis, some of the basic states of consciousness of the vacuum can be associated in a local view to the known elementary particles (leptons and quarks, which together constitute the so-called fermions) and to the fundamental interactions, also mediated by elementary particles (called bosons ). In quantum physics, at the present state of knowledge of elementary particles, we have 6 quarks and 6 leptons (these are all fermions), divided into three generations of up and down, with their antimatter counterparts; 8 gluons which are the mediators of strong interactions, the photon which is the mediator of electromagnetic interactions, three bosons which mediate weak interactions; we also have the Higgs boson, and some hypothetical gravitons that would be the mediators of the gravitational interaction (for which, however, a satisfactory quantum theory does not yet exist). A total of 24 fermions, plus 12 bosons mediators of interactions: 36 particles consisting of fermions and bosons. Moreover the Higgs boson should be added. 36 are also the most stable basic states of consciousness, which I would associate with the fermions and interaction bosons found experimentally in large accelerators. Of the other 28 elements of consciousness that are more difficult to activate at the local level, one I would associate with the Higgs boson, another with the graviton. The remainder constitute elements of consciousness without a counterpart in terms of wave or particle in known local observation.

These elements of a purely non-local character, and as such not attributable in any way to elementary particles observable in local experiments, could be the constituents of dark matter and dark energy, whose presence is recognized by indirect measurements of a cosmological nature. In this case, the search for dark matter and dark energy in the large astrophysical experiments that are currently underway should change the investigation objective, focusing on the non-local effects; these effects are verifiable in experiments conceptually different from the classical ones of direct observation, to which I will mention later.

We are in the field of pure speculation, in which I prefer not to go further, also because I do not have a specific competence in elementary particle physics. However, the hypothesis of a primary consciousness as the fundamental substratum of the universe does not seem unreasonable to me, on the contrary it appears perfectly consistent with the synchronic principle that manifests itself both in the recent developments of quantum physics, and indirectly through the probabilistic considerations of an unguided random evolution. It is therefore natural to hypothesize that the information exchanged by the agents of consciousness, self-defined according to a criterion of free choice and non-random evolution, is that of quantum information.

In the next chapter I will try to draw guidelines to understand how the elements of consciousness spontaneously evolve towards a broader awareness; furthermore I will indicate a principle methodology for verifiability with the experiment of the new non-local predictions related to the synchronic principle in the simplest cases.

## 4. The evolution of consciousness

After our definition of the elementary agents of consciousness of the vacuum state as coherent superpositions of hexagrams consisting of arrays of 6 quantum bits, it remains for us to indicate how evolution takes place starting from these simple elements.

In our model, evolution occurs by not randomly determined attempts, since the elements of consciousness have a possibility of choice, limited to the minimum in the elementary case but sufficient to guide the evolution of the Universe towards the conditions necessary for the development of biological organisms. During the process of evolution, consciousness becomes more and more complex through the synchronic aggregation of elementary agents, and the possibility of choice gradually increases until it reaches the complexity of human consciousness, which we know well from direct self-referential experience, and perhaps other forms of very evolved conscience that we still do not have the possibility to perceive with our level of conscience. Our intelligence allows us many free and responsible choices, so it is now evident that our evolution does not occur by attempts determined (only) by chance but is partly guided by awareness.

The dynamics of the Universe allows us to define a time, whose arrow indicates the direction in which the evolution of consciousness naturally occurs. In our local view time is a primary concept, fundamental to define scientific laws but in a synchronic view time is a secondary concept, not intrinsic to the Universe but rather it represents the way in which our mind is able to sequentially conceive the evolutionary process. In the local paradigm, space-time is the basic substratum on which the laws of nature act through which we recognize a regularity in the dynamic process of evolution.

In the local view, scientific laws appear to us as immutable and universal; they are part of the intrinsic nature of the Universe that moves as a result of these laws expressed by deterministic mathematical equations. This is the causal paradigm of evolution at the basis of the classical scientific method, in which consciousness is a random "accident", which could or could not happen and which indeed had an almost zero probability of emerging: a true miracle of combinations starting from the first moments of the birth of the Universe. Life has come to us through an incredible combination of circumstances that a mathematical calculation would have no chance of happening synchronously.

In a synchronic model, awareness is primary and exists before the appearance of the space-time interface created by the level of consciousness of the human mind: the thrust of evolution in the synchronic manifestation of the Universe is the tendency of awareness to evolve towards forms increasingly inclusive, through attempts that are partly casual and partly guided in such a way that not all possibilities are equally likely; this conscious guide has preferably selected those conditions which favor life and the development of consciousness.

This synchronicity manifests itself at the level of our local vision as a regularity and harmony that we interpret as scientific laws that operate on matter in space-time.

So in a synchronic paradigm we must admit the priority presence of awareness - even before space-time - and inseparable from the state of emptiness. Evolution is determined by universal awareness, of which we participate together with everything else that exists.

The basic consciousness agents of the quantum vacuum have a very limited choice, the minimum possible, which led to the definition of the 64 hexagrams, corresponding to arrays of 6 quantum bits. In the inclusive process of the evolution of consciousness from the state of emptiness, we must admit multiple possibilities of choice for the most advanced forms of consciousness, some of which are not independent but conditioned. In the quantum model, conditional probabilities must also exist in coherent superposition, leading to a formidable mathematical complication. Fortunately, we have the progress made in the field of quantum information to be able to draw the guidelines.

The definition of elementary particles as conscious agents characterized by qu-bit arrays could allow us to predict interaction effects not yet seen in collisions between particles in large accelerators, through the use of quantum information processing to determine the correlation. This could be a new field of research in physics applied to elementary particles.

The field of investigation is completely new, but it could still make use of the mathematical considerations developed to treat symmetries in quantum physics and apply them to non-local effects. The quantum computer itself could be the system on which increasingly complex synchronic experiments are carried out, compatibly with the development of quantum technologies.

Many technologies are already present, and it would be enough to shift the focus of the experiment to synchronic phenomena to make much progress in the development of synchronic science. What we normally do in experiments according to the classical scientific method is to observe the effect that a local action causes in the same part of the system in which the action is physically carried, disinterested in what happens in the rest of the collective system.

Synchronicity experiments will have a strong non-local component but also a locally observable relapse. We take for this purpose an entangled system. We divide it into two parts, so that there cannot be a cause-effect correlation between the two, and to carry out local actions on one part and verify the changes of state of the second part on which we do not exercise any action. Although we do not interact with the part that serves as verification, changes of state will occur due to the synchronic link with the part we are interacting with directly. These changes of state, compared with the predictions obtainable in a synchronic theory, can constitute experimental verifications.

In the simplest cases, information can be obtained even by using simple considerations of quantum physics. For example, we can think of experiments with elementary particles, where the processing of information could be simple enough to be foreseen even without the support of a quantum computer. For example, the experiment conducted at Stanford by the "BABAR" collaboration (Lees 2012) to directly determine a temporal asymmetry (called violation of T) in the decay of neutral mesons $B^0$, is based on the measurement of the probability of transformation between pairs of correlated quantum states made up of matter and antimatter: since the pairs of mesons are created in a state of entanglement, the two mesons, while moving away from each other, remain intimately connected without assuming a well-defined identity. In other words, both are simultaneously both the particle and the antiparticle in a quantum superposition state. Due to entanglement, when the one of the two mesons decays spontaneously, the identity of the other meson of the pair is also determined at the same time. This property becomes a formidable scientific investigation tool: you can select the state of the second meson without having to observe it, by choosing a particular decay of the first meson. This property was used in the experiment to determine the violation of T.

This is a quantum physics experiment, in which synchronicity was only a tool and not the object of the investigation; but if we analyze the procedure used, it contains in elementary form the characteristics that the more complex synchronic experiments must have: by applying a causal action to a part of the correlated system - in this case it was the simple observation of the spontaneous decay of the first boson of the pair - information is obtained and a transformation is generated in the other part of the system, which is the real object of the investigation. Since the state is one of superposition, the change that occurs in the second part cannot be predicted in a deterministic way; in this case we had a 50% chance of dealing with matter and another 50% with antimatter.

This is what happens in every experiment in synchronic science: in any case we will not have the certainty of obtaining a desired effect locally, but we can try to maximize the probability of manifestation of the desired effect, and select only the cases of interest.

## 5. Conclusions

We are accustomed to considering scientific laws as immutable natural laws, valid in any place and at any time. Indeed this belief was assumed as the principle of symmetry of the universe, and associated with two fundamental conservation laws in both classical and quantum physics: that of energy-momentum.

Furthermore, the physical laws are assumed to be the same for each observer, and this belief is the basis of Einstein's theory of relativity. Another consolidated belief is that reality is unique, that is, that nature cannot manifest itself through parallel realities.

In a synchronic view we must revolutionize these beliefs that appear so natural and obvious to us. First of all we have seen that space and time are secondary concepts, created by the human mind to conceive the synchronic reality of the Universe in a sequential way; therefore there are no immutable laws in space-time except in the representation of human consciousness. We have also seen that the human representation of material reality is only an interface that simplifies the synchronic complexity of the universe; the function of this interface is not to create a true model of reality, but to create symbolic icons through which to quickly and efficiently manage interactions with "our world" for the purposes of survival and natural selection, toward an increasingly inclusive consciousness. The primary reality in the universe is awareness that self-organizes itself into elements of consciousness of ever more inclusive complexity: from the fundamental elements of emptiness and matter, up to the most advanced forms of consciousness. The most inclusive forms of conscience are aware of themselves and of the less evolved ones, while they are unable to have direct experience of the most evolved and non-local forms of consciousness.

Each level of consciousness therefore has its own reality, its world described by its laws: there are as many parallel realities as there are levels of inclusion of consciousness.

In a synchronic universe, in which consciousness and choice are primary aspects, it is not possible to establish principles or physical laws valid at every level of inclusion. Physical laws and universal principles can be recognized at the lowest level, that of vacuum and elementary particles (and in general the level of inert matter).

At the first level, the material one, the degree of freedom of choice is low and is limited to the choice of universal constants that characterize elementary particles and fundamental interactions in order to allow the conditions for the emergence of life and more inclusive consciences. This is the level that all conscious agents share and on which everyone can agree. Given the scarce freedom of choice at this level it is possible to make predictions described by exact physical laws: this is the world described by Quantum Physics, where experiments can be carried out whose results are comparable with theoretical predictions.

The emergence of synchronicity from the quantum model however warns us that this is only the initial level of consciousness, and a reductionist vision of reality is generally not possible.

The reductionist model of reality is very reassuring for the rational mind, which would like to be able to predict events with deterministic certainty and fear the unexpected. This defensive attitude has tended to relegate scientific considerations to those aspects of reality that are simple enough to be somehow tractable with the classical scientific method, considering as unscientific those phenomena that are not rationally accessible starting from simple schematizations. We dramatized, or even avoided by branding them as superstition, all those phenomena in which non-locality constitutes an inevitable structural element, exasperating the contrast between what is "scientific" and what cannot be understood in a reductionist model. This attitude has dried up our conscience, and restricted the field of priority interest to the immediate material and economic aspects. The world that is accessible to our perception is much wider, and this text is an invitation to recover the potential of the human level of consciousness also from a scientific point of view.

The level of consciousness immediately above the material one is that of quantum synchronicity. At this level we have to give up making predictions through sequential logic, and we have to think in overall holistic terms. This is the level that all biological organisms share, including ecosystems and the "living planet", our earth. Physical laws at this level cannot be deterministic and reproducible, since each organism is unique and irreplaceable.

Each agent of consciousness has its reality and its laws that describe its world, with which it is synchronously connected, inseparable from it.

If we assume that the mathematical formalism of Quantum Physics continues to apply, it can be used as a guide tool for the intuitions to be verified in direct experience: in particular, the quantum computer could process information without reducing it to the classical one and help human consciousness to make the right choices, that is, the most inclusive ones, in all situations that can be programmed through quantum algorithms. This use of the quantum computer could lead to experimentally verifiable situations, the value of which could be shared by the scientific community, despite the impossibility of absolute reproducibility.

These verifications could be the driving force for a transformation in a more inclusive sense of human consciousness as a whole.

# References

Arute, K et al. 2019, "Quantum supremacy using a programmable superconducting processor", Nature Vol. 574, p505.

Aspect, A, Grangier, P & Roger, G 1981 "Experimental Test of Realistic Local Theories via Bell's Theorem", Phys. Rev. Lett. Vol.47, p.460.

Bohm, David 1980, "Wholeness and the implicate order", Routledge - London and New York.

Caldeira, AO & Leggett, AJ 1981, "Influence of dissipation on quantum tunneling in macroscopic systems", Phys. Rev. Lett. Vol.46, pp211–214.

Chiorescu, I, Nakamura, Y, Harmans, CJPM, & Mooij, JE 2003, "Coherent quantum dynamics of a superconducting flux qu-bit", Science Vol.299, p1869.

Emary, C, Lambert, N & Nori, F 2014, "Leggett Garg inequalities", Rep. Prog. Phys. Vol.77, 016001.

Freedman, SJ & Clauser, J 1972, "Experimental test of local hidden-variables theories", Phys. Rev. Lett. Vol.28, p938;

Fields, C, Hoffman, DD, Prakash C, & Singh, M 2018 "Conscious agent networks: formal analysis and application to cognition", Cognitive Systems Research Vol.47, p186.

Fields, Chris 2016, "De-compositional equivalence: a fundamental symmetry underlying quantum theory" Axiomates Vol.26, p279.

Giustina, M et al. 2015, "Significant-Loophole-free Test of Bell's Theorem with entangled photons" Phys. Rev. Lett. Vol.115, p250401.

Groeblaker et al. in 2007, "An experimental test of non-local realism", Nature Vol.446, p871.

Hansen, B et al. 2015, "Loophole-free Bell inequalities violation using electron spins separated by 1.3 kilometers", Nature Vol.526, p682.

Hoffman, DD "Conscious Realism and the Mind-Body Problem", Mind & Matter Vol.6, pp 87-121 (2008);

Knee, GC et al. 2016, "A strict experimental test of macroscopic realism in a superconducting flux qubit", Nature Communication Vol.7, p13253.

Leggett, AJ, Ruggiero, B & P. Silvestrini 2004, "Quantum Computing and Quantum Bits in Mesoscopic Systems", Kluwer Academic/ Plenum Publisher.

Lees J.P et al. (The BABAR Collaboration) 2012, "Observation of Time-Reversal Violation in the $B^0$ Meson System " Phys. Rev. Lett. Vol.**109**, p211801.

Martinis, JM, Devoret, MH & Clarke, J 1987, "Experimental tests for the quantum behavior of a macroscopic degree of freedom: the phase difference across a Josephson junction", Phys. Rev. B35, p4682.

Nakamura, Y, Chen, CD, & Tsai, JS 1997, "Spectroscopy of energy level splitting between two macroscopic quantum states of charge coherently superposed by Josephson coupling", Phys. Rev. Lett. Vol79, pp2328.

Nielsen, Michael & Chuang, Isaac 2000 "Quantum Computation and Quantum Information", Cambridge University Press
-2010 (second edition).

Rouse, R, Han, S, & Lukens, JE, 1995, "Observation of resonant tunneling between macroscopically distinct quantum levels", Phys. Rev. Lett. Vol.75, p1614.

Silvestrini, P, Palmieri, VG, Ruggiero, B, & Russo, M, 1997, "Observation of energy level quantization in under-damped Josephson junctions above the classical-quantum regime crossover temperature", Phys. Rev. Lett. Vol.79, pp3046–3049.

Wan, C et al. 2016, "A Schroedinger cat living in two boxes", Science Vol.352, p1087.

Zurek, WH 1981, "Pointer basis of a quantum apparatus: into what mixture does the wave packet collapse?" Phys. Rev. D24, p1516.

Zurek, WH 2003, "De-coherence, einselection and the quantum origin of the classical" Reviews of Modern Physics Vol.75 (3), p715.